\documentclass[preprint]{JHEP3}
\usepackage{epsfig}
\usepackage{graphicx,psfrag}
\usepackage{dcolumn}
\usepackage{bm}
\usepackage{amsmath,amssymb}

\newcommand{\beq}{\begin{equation}} 
\newcommand{\eeq}{\end{equation}}   
\newcommand{\bea}{\begin{eqnarray}} 
\newcommand{\eea}{\end{eqnarray}} 
\newcommand{\hf} {\frac{1}{2}} 
              
\newcommand{\nn}{\nonumber\\}

\newcommand\eqn[1]{Eq.\,(\ref{#1})}

\newcommand\fig[1]{Fig.\,{\ref{#1}}}

\def\tu{{\tilde u}} 
\def\tv{{\tilde v}} 

\def\tV{{\tilde V}}

\def\mr#1{{\mathrm{#1}}} 
 
\def\eq#1{(\ref{#1})} 

\title{Phase Structure and Compactness}

\author{I. N\'andori$^a$, S. Nagy$^b$, K. Sailer$^b$, A. Trombettoni$^c$\\
$^a$Institute of Nuclear Research, P.O.Box 51, H-4001 Debrecen, Hungary\\
$^b$Department of Theoretical Physics, University of Debrecen, Debrecen, Hungary\\
$^c$SISSA and INFN, Sezione di Trieste, via Bonomea 265, 
I-34136 Trieste, Italy}

\abstract{
In order to study the influence of compactness on low-energy properties, 
we compare the phase structures of the compact and non-compact 
two-dimensional multi-frequency sine-Gordon models. It is shown that the 
high-energy scaling of the compact and non-compact models coincides, 
but their low-energy behaviors differ. The critical frequency $\beta^2=8\pi$ 
at which the sine-Gordon model undergoes a topological phase transition 
is found to be unaffected by the compactness of the field since it is determined 
by high-energy scaling laws. However, the compact two-frequency sine-Gordon 
model has first and second order phase transitions determined by the low-energy 
scaling: we show that these are absent in the non-compact model.} 


\preprint{}

\begin{document}

\section{Introduction} 
\label{intro}
          
The computation in a completely controlled way of renormalization 
group (RG) flows in gauge theories is at date a challenging issue.
A major reason for such difficulties is the fact that, one must adopt a 
regularization scheme which incorporates a gauge invariant cutoff 
even for approximated treatments of exact RG equations. A related 
problem is the determination of critical properties and phases of compact 
field theories, since, compactness can be considered as one of the simplest 
realization of the gauge symmetry \cite{zinn} and a general treatment for the 
controlled computation of renormalization flows in gauge theories would 
apply to the RG study of compact field theories.

{\em Sic stantibus rebus}, in order to study the influence of compactness on 
low-energy properties, it would be then relevant to compare the phase 
structure of a field theory with the fields being respectively compact and 
non-compact. In this paper we perform such comparison for the multi-frequency 
sine-Gordon (SG) model in $1+1$ dimensions. Several reasons lead us to the 
choice and use of such a model for the purpose of studying the effects of 
compactness: first, the single-frequency SG is a paradigmatical example of 
integrable field theory \cite{GMbook}, very well studied in the last four decades. 
Second, the rich phase structure of the compact double-frequency SG has been 
the subject of intense study 
\cite{DeMu1998,FGN,BaEtAl2000,BaEtAl2001,MuRiSo,multi_sg,TaWa2006}, 
and, last but not least, the non-compact multi-frequency SG model (MFSG) 
can be studied using non-perturbative RG methods \cite{rg}.
 
The importance of the SG model stems from the fact that it is directly related 
to interacting fermionic field theories through bosonization \cite{gogo}. In low 
dimensions exact bosonization rules enable one to reformulate fermionic and 
gauge models in terms of elementary scalar fields. The equivalence between 
the massive Thirring model and the sine-Gordon (SG) scalar theory \cite{sg_model} 
is a well-known example. Two-dimensional gauge models like the multi-flavor 
quantum electrodynamics (QED$_2$) \cite{lsg_model,qed2_qcd2,qed2}
and the single-flavor quantum chromodynamics (QCD$_2$) \cite{single_qcd2}
can also be rewritten as a multi-component SG theory where the SG fields  are 
coupled by a mass-matrix. It has been also shown that various aspects of the 
low-energy QCD$_2$ with multi-flavors (and with unequal quark masses) 
can be described by the so-called generalized SG model \cite{multi_qcd2} of 
which reduced sub-model is the MFSG model with non-compact field variable.

Moreover, the SG model, the simplest non-trivial quantum field theory which can 
be used to study confinement phenomena, has already received a considerable 
attention in several areas of physics. For example, in string theory the SG model 
is assumed to be related to the classical string on specific manifolds  
\cite{sg_classical_string} and the possible contribution of new type of SG models 
to brane profile has also been studied \cite{sg_string} in the framework of 
Randall-Sudrum \cite{rs} theory. SG type model has recently been investigated in 
3+1 dimensions in the context of axion physics \cite{sg_axion}. Furthermore, the 
SG model is used as a textbook example for integrable systems and it has many 
applications in condensed matter and statistical physics as well, e.g. coupled SG 
models were successfully used to describe the vortex dynamics of layered high 
transition temperature superconductors \cite{NaEtAl2007}. Another attractive 
property of low-dimensional SG models is that they provide us an excellent 
playground to test and compare  various types of non-perturbative methods 
\cite{Abdalla}. For example, SG type models have already been investigated in 
the framework of Integrable and Conformal Field Theory (CFT) \cite{YuZa} and 
the exact functional RG treatment for these periodic models has also been 
developed \cite{sg,NaEtAl2007,qed2,qed2_qcd2,sg_prl}.

Our goal in this paper is to consider the non-compact MFSG model by means 
of the functional RG approach and to compare our findings to those obtained by 
other methods for the non-compact and compact models, as well. In particular, 
we investigate the influence of the compact or non-compact nature of the field 
variable on the low-energy behavior of the MFSG model whose action reads as 
\cite{DeMu1998,FGN,BaEtAl2000,BaEtAl2001,MuRiSo,multi_sg,TaWa2006}
\beq
\label{mfsg}
{\cal S}_{\rm MFSG} = \int \mathrm{d}^2 x
\left[
\frac{1}{2} \partial_{\mu}\phi \partial^{\mu}\phi
- \sum_{i}^{n} \mu_{i} \cos(\beta_{i} \phi + \delta_{i}) 
\right]
\eeq
which contains $n$ cosine terms where $\phi$ is a real scalar field, 
$\beta_{i}\in\mathbb{R}$ are the frequencies, $\beta_{i}\ne\beta_{j}$
if $i\ne j$, $\mu_{i}$ are the coupling constants (of dimension
mass$^{2}$ at the classical level) and $\delta_{i}\in\mathbb{R}$
are the phases in the terms of the potential. Let us note that the
MFSG model is usually defined on the two-dimensional Minkowski 
space, however, in this paper we use the Euclidean action which 
is more convenient  for an RG study and it is generally assumed to 
be suitable for mapping out the phase structure of the model.

Two cases can be distinguished according to the periodicity properties
of the model. The first one is the rational case, when the potential is a 
trigonometric function: the ratios of the frequencies $\beta_{i}$ are 
rational and consequently, the potential is periodic. Let the period of the 
potential be $2\pi \beta$ in this case. Then the target space of the field $\phi$ 
can be compactified: $\phi \equiv\phi+2k \beta \pi,$ where $k\in\mathbb{N}$ 
can be chosen arbitrarily. The model obtained in this way is called the 
$k$-folded multi-frequency SG model \cite{BaEtAl2000}. The other case is 
the irrational one, when the potential is not periodic. 
We restrict our attention to the rational case in the present paper. 

At the quantum level the theory can be considered as a perturbation of 
its high-energy/ultraviolet (UV) limiting conformal field theory 
\cite{DeMu1998,FGN,BaEtAl2000,BaEtAl2001,MuRiSo,multi_sg,TaWa2006}
\beq 
\label{conform}
{\cal S}_{\rm MFSG} = {\cal S}_{\rm CFT} + {\cal S}_{\rm pert},
\eeq
where
\begin{align}
&{\cal S}_{\rm CFT} = 
\int \mathrm{d}^2 x \frac{1}{2} \partial_{\mu}\phi \partial^{\mu}\phi, 
\nn
&{\cal S}_{\rm pert} =-\frac{1}{2} \int \mathrm{d}^2 x 
\sum_{i=1}^{n}(\mu_{i}e^{i\delta_{i}}V_{\beta_{i}}
+\mu_{i}e^{-i\delta_{i}}V_{-\beta_{i}}),
\nonumber
\end{align}
with the vertex operator $V_{\omega}=:e^{i\omega\phi}:$ which  
corresponds to a primary field with conformal dimensions 
$\Delta_{\omega}^{\pm}=\Delta_{\omega}=\frac{\omega^{2}}{8\pi}$
in the UV limit and the upper index $\pm$ corresponds to the left/right 
conformal algebra and : : denotes the conformal normal ordering. 
Correspondingly, the dimensions of the couplings in the UV limit at the 
quantum level are $[\mu_{i}]=({\rm mass})^{2-2\Delta_{i}}$ with 
$\Delta_{i}\equiv\Delta_{\beta_{i}}$.

It was shown by semiclassical (mean-field/Landau-Ginzburg) analysis  
\cite{DeMu1998} and by means of form factor perturbation and truncated 
conformal space approaches  \cite{DeMu1998,BaEtAl2001,multi_sg,TaWa2006}
that (first and second order) phase transitions occur in the compact MFSG 
model as the coupling constants are tuned appropriately (assuming that 
$n>1$). For example, according to the semi-classical results \cite{DeMu1998}, 
the double-frequency SG model (for $\delta_1=0$, $\delta_2 =\pi/2$ and 
$\beta_2= \beta_1/2$)
\beq
\label{dfsg}
{V}_{\rm DFSG}(\phi) = 
- \mu_1 \cos(\beta_1 \phi) +  \mu_2 \sin\left(\frac{\beta_1}{2} \phi\right) 
\eeq
undergoes a second order (Ising-type) phase transition at $\mu_2 = 4\mu_1$
\cite{DeMu1998}. This second order phase transition was found to appear for 
all frequencies $0<\beta^2<8\pi$ beyond the semi-classical level, see e.g. the 
phase diagram in Fig. 7.5 of \cite{BaEtAl2001} which was determined by form 
factor perturbation theory and truncated conformal space approach. The Ising-type 
phase transition was also confirmed by renormalization group techniques based on 
operator product expansion in real space  \cite{rg_multi_sg}.  Let us note, that in 
this case the field variable is defined as a compact variable.  It was also argued 
that the MFSG model reduces to the classical (single-frequency) SG model in the 
limit of $\delta_i \to 0$ for $i=1,2, ...,\infty$ and $\mu_i \to 0$ for $i=2,3, ...,\infty$. 
It was also shown that the SG model defined by the action which contains 
non-compact field variable \cite{sg,sg_prl} belongs to the universality class 
of the two-dimensional Coulomb gas and the two-dimensional XY model, 
consequently, its phase transition at $\beta_c^2 = 8 \pi$ is a topological or 
Kosterlitz-Thouless-Berezinskii (KTB) type one \cite{KTB}. It is also known 
that the classical SG model with a compact field variable also possesses a 
topological phase transition at $\beta^2_c =8\pi$. Therefore, on the one hand, 
the MFSG model with compact field variable has Ising-type phase transitions 
(for $n>1$), on the other hand in case of a single cosine (for $n=1$) with 
compact and non-compact fields the model has a topological phase transition. 
Consequently, it represents an excellent toy model to study the influence of 
the compactness on the phase structure and the low-energy behavior of the 
model.

Our goal is to compare the UV/IR scaling behavior and the phase structure 
of the MFSG models with compact and non-compact field variables: 
we study the MFSG theory with non-compact fields by means of the functional 
RG method in the local potential approximation, discussing the comparison 
with the available results for the compact model 
\cite{DeMu1998,FGN,BaEtAl2000,BaEtAl2001,MuRiSo,multi_sg,TaWa2006,rg_multi_sg}. 
The structure of our paper is the following: a 
brief introduction of RG equations 
used for the renormalization of the non-compact MFSG model is given in 
Section \ref{rg}. In Section \ref{symmetry}, the connection between RG 
equations and symmetries of the MFSG model is discussed. The UV and IR 
scaling laws of the non-compact MFSG model are determined and compared 
to those of the compact model in Sections \ref{uv} and \ref{ir}, respectively.
Finally, Section \ref{sum} presents the summary and our concluding remarks.

\section{Renormalization Group Approach} 
\label{rg}
In this section we briefly discuss the functional RG  equations used for the
renormalization of the MFSG model. The differential RG transformations are 
realized via a blocking construction \cite{Wi1971}, the successive elimination 
of the degrees of freedom which lie above the running UV momentum cutoff 
$k$. Consequently, the effective theory described by the blocked action contains 
quantum fluctuations whose frequencies are smaller than the momentum cutoff. 
This procedure generates the functional RG flow equation 
\cite{RiWe1990,We1993,Mo1994} 
\beq
k \partial_k \Gamma_k [\phi] = \hf \mathrm{Tr}  
\left( \Gamma^{(2)}_k [\phi] + R_k \right)^{-1}  
k \partial_k R_k 
\nonumber
\eeq 
for the effective action $\Gamma_k [\phi]$ when various types of regulator 
functions $R_k$ are used, where $\Gamma^{(2)}_k [\phi]$ denotes the 
second functional derivative of the effective action (see e.g. \cite{rg}). 
Here $R_k$ is a properly chosen IR regulator function which 
fulfils a few basic constraints to ensure that $\Gamma_k$ approaches the 
bare action in the UV limit ($k\to\Lambda$) and the full quantum effective 
action in the IR limit ($k\to 0$). Indeed, various renormalization schemes 
are constructed in such a manner that the RG flow starts at the bare action 
and provides the effective action in the IR limit, so that the physical predictions 
(e.g. fixed points and critical exponents) are independent of the renormalization 
scheme particularly used \cite{scheme,sg_scheme,litim_scheme}.

Since RG equations are functional partial differential equations it is not 
possible to solve them in general, hence, approximations are required. 
One of the commonly used systematic approximation is the truncated 
derivative expansion where the effective action is expanded in powers 
of the derivative of the field \cite{scheme,sg_scheme,litim_scheme},
\beq
\Gamma_k [\phi] = \int_x \left[V_k(\phi) 
+ Z_k(\phi) \hf (\partial_{\mu} \phi)^2 + ... \right].  
\nonumber
\eeq 
In the local potential approximation (LPA) higher derivative terms are 
neglected and the wave-function renormalization is set equal to constant, 
i.e. $Z_k \equiv 1$.  In this paper we use two types of RG equations  (i.e. two 
different IR regulators $R_k$), namely  the Wegner-Houghton \cite{WeHo1973} 
and the Polchinski \cite{Po1984} RG approaches. However, let us note that in 
the LPA, the two-dimensional Wegner-Houghton RG equation is mathematically 
equivalent (see e.g. \cite{sg_scheme}) to the effective average action RG 
equation \cite{RiWe1990,We1993} with the power-law regulator 
$R_k(p^2)\equiv p^2 (p^2/k^2)^{-b}$ \cite{Mo1994} with $b=1$ and the 
functional Callan-Symanzik RG equation \cite{internal}.

\subsection{Wegner--Houghton, effective average action and functional 
Callan-Symanzik  RG equations} 
In this section, we consider three types of RG equations, namely the Wegner--Houghton, 
the effective average action with power-law regulator ($b=1$) and the functional 
Callan-Symanzik RG equations which have the same form in LPA for $d=2$ 
dimensions \cite{sg_scheme}.

The blocking in momentum space, i.e. the integration over the field 
fluctuations with momenta of the magnitude between the UV scale $\Lambda$ 
and zero is performed in successive blocking  steps over infinitesimal 
momentum intervals $k\to k-\Delta k$ each of which consists of the splitting 
the field variable, $\phi=\varphi+\phi'$ in such a manner that $\varphi$ and 
$\phi'$ contain Fourier modes with $|p|<k-\Delta k$ and $k-\Delta k<|p|<k$, 
respectively and the integration over $\phi'$ leads to the Wegner--Houghton 
(WH) RG equation \cite{WeHo1973}  
\beq 
\label{WHdim} 
\left(2 + k\partial_k \right) {\tilde V}_k ( \phi) =  
-\frac1{4\pi}\ln\left(1 + {\tilde V}''_k (\phi) \right) 
\eeq 
with $\tilde V''_k(\phi) = \partial^2_{\phi} \tilde V_k(\phi)$ for the 
dimensionless local potential ${\tilde V}_k = k^{-2} V_k$ for $d=2$ 
dimensions in the leading order of the derivative expansion, in the LPA 
when $\phi$ reduces to a constant. (Below we suppress the notation of 
the field-dependence of the local potential and use notations with tilde 
for dimensionless quantities where the dimension is taken away by the 
appropriate power of the gliding cutoff $k$.) The differentiation with 
respect to the field variable and the multiplication with $1+\tV_k''$ 
leads to the derivative form of the WH--RG equation \cite{sg_scheme}
\beq\label{derwh} 
(2+k\partial_k)\tilde V'_k =  
-\tilde V''_k(2+k\partial_k)\tilde V'_k-\frac1{4\pi}\tilde V'''_k. 
\eeq 
This equation is obtained by assuming the absence of instabilities for 
the modes around the gliding cutoff $k$. The WH-RG scheme which uses the 
sharp gliding  cutoff $k$ can also account for the spinodal instability, 
which appears when the restoring force acting on the field fluctuations 
to be eliminated vanishes, $1+\tilde V''_k(\phi)=0$ at some finite scale 
$k_{\mr{SI}}$  and the resulting condensate generates tree-level contributions 
to the evolution equation. The saddle point $\phi'_0$ for the single  
blocking step $k\to k-\Delta k$ is obtained by minimizing the action,  
$S_{k-\Delta k}[\phi]=\min_{\phi'_0}\left(S_k[\phi + \phi_0']\right)$. 
The restriction of the space of saddle-point configurations to that of 
the plane waves  $\phi'_0 = \rho \cos(k_1 x)$  gives \cite{tree}
\begin{align} 
\label{treedim} 
{\tilde V}_{k-\Delta k}(\phi) = \min_\rho \left[\rho^2 +\hf 
\int_{-1}^1 du {\tilde V}_k(\phi + 2\rho \cos(\pi u)) \right] 
\end{align} 
in LPA, where the minimum is sought for the amplitude $\rho$ only. It 
was shown that the tree-level RG equation \eq{treedim} leads to the local  
potential \cite{sg_scheme}
\begin{equation}
\label{treewhpot}
\tilde V_{k\to 0} = -\hf \phi^2 + c \phi + {\rm const},
\end{equation}
which can also be obtained as the solution of $1+\tilde V''_{k\to 0}(\phi)=0$
(in case of a $\phi \to -\phi$ symmetry the linear term vanishes,  $c=0$).
Therefore, if SI occurs during the RG flow at some scale $k_{\mr{SI}}>0$, 
then Eqs. \eq{WHdim}  or  \eq{derwh} should be applied only for scales 
$k>k_{\mr{SI}}$, and the tree-level renormalization Eq.\eq{treedim} or
Eq.\eq{treewhpot} should be performed at scales $k<k_{\mr{SI}}$.

For $d=2$ dimensions the effective average action (EAA) RG equation
with power-law regulator can be written in the LPA as  
\beq
\label{power_effective_rg} 
(2+k \partial_k) {\tilde V}_k = - \frac{1}{4\pi}  
\int_{0}^{\Lambda^2/k^2} \mathrm{d}y 
\frac{(-b) y^{-b} \, y}{y(1+y^{-b}) +{\tilde V''}_k}
\eeq 
with $y = p^2/k^2$. For arbitrary parameter value $b$, the propagator on the 
right hand side of Eq. \eq{power_effective_rg} may develop a pole at some 
scale $k_{\mr{SI}}$ and at some value of the field $\phi$  for which 
$\tilde V_k''(\phi) = -C(b) = - b/(b-1)^{(b-1)/b}$ holds, which signals the 
occurring of SI. The infrared singularity of the functional RG equation is 
supposed to be related to the convexity of the effective action for theories 
within a phase of spontaneous symmetry breaking \cite{We1993}. It was shown 
that in such a case one has to seek the local potential for $k<k_{\mr{SI}}$ by 
minimizing $\Gamma_k$ in the subspace of inhomogeneous (soliton like) field 
configurations and ends up with the result \cite{We1993,rg} 
\begin{equation}
\label{eaafppot}
\tilde V_{k\to 0} = -\hf C(b)\phi^2 + c \phi + {\rm const}.
\end{equation}
It is worthwhile noticing that Eq. \eq{power_effective_rg} with the power-law 
regulator leads to the WH-RG equation \eq{WHdim}, and \eqn{eaafppot} 
leads to \eqn{treewhpot} for $b=1$ as well as for $b\to \infty$ in the limit 
$\Lambda\to \infty$. This feature holds only for $d=2$.

In the functional Callan-Symanzik (CS) type internal space RG method 
\cite{internal}, the successive elimination of the field fluctuations is performed 
in the space of the field variable (internal space) as opposed to the usual RG 
methods where the blocking transformations are realized in either the 
momentum or the real (external) space. The functional CS--RG equation 
for the one-component scalar field theory for dimensions $d=2$ in the LPA reads 
\bea  
\label{internal_rg}  
&(2+\lambda \, \partial_{\lambda}) \,\, \tilde V_{\lambda} =  
- \frac{1}{4\pi}  
\ln \left(1 + \tilde V''_{\lambda} \right) 
\eea  
with the control parameter $\lambda$.
This equation is mathematically equivalent to the two-dimensional 
WH--RG equation in the LPA assuming the equivalence of the scales 
$\lambda \equiv k$. However, for dimensions $d\neq 2$ the functional
CS--RG and the WH--RG differ from each other. Assuming 
the above mentioned equivalence of the scales $\lambda$ and $k$, there 
occurs the same singularity in the right hand side of \eq{internal_rg} as 
the one in the WH-RG approach. Therefore, the functional CS--RG  signals 
the SI with the vanishing of the argument of the logarithm in the right hand 
side of \eq{internal_rg}. The solution of \eq{internal_rg} provides the scaling 
laws down to the scale $k_{\mr{SI}}$ and one has to turn to the tree-level 
renormalization with the help of the WH-RG approach in order to determine 
the IR scaling laws. 

Since the WH--RG, EAA--RG with power-law regulator ($b=1$) and the 
functional CS--RG equations have the same form in LPA for $d=2$ 
dimensions, in this paper we refer to them as the WH--RG equation.

\subsection{Polchinski's RG equation} 
In Polchinski's RG (P--RG) method \cite{Po1984} the realization of the 
differential RG transformations is based on a non-linear generalization of 
the blocking procedure using a smooth momentum cutoff. In the infinitesimal 
blocking step the field variable $\phi$ is split again into the sum of a 
slowly oscillating IR and a fast oscillating UV components, but both fields  
contain now low- and high-frequency modes, as well, due to the smoothness of 
the cutoff. Above the moving momentum scale $k$ the propagator for the IR 
component is suppressed  by a properly chosen smooth regulator function $K(y)$ 
with $y=p^2/k^2$, $K(y)\to 0$ if $y>>1$, and $K(y)\to 1$ if $y<<1$. The P--RG 
equation in LPA for $d=2$ dimensions reads as 
\beq
\label{polch} 
(2+k\partial_k)\tilde V_k = -[\tilde V'_k]^2 K'_0+\tilde V''_k I_2, 
\eeq 
where $K'=\partial_y K(y)$, $K'_0 = \partial_{y} K(y)\vert_{y=0}$ and 
$I_2= (1/4\pi)\int_0^\infty dy K'(y) = - 1/4\pi$. The parameter $K'_0$ 
can be eliminated by the rescaling of the potential and the field variable, 
consequently, it does not influence the physics. In order to make the 
comparison of the RG flows obtained by various RG methods straightforward, 
we choose $K_0^\prime=-1$ for which the linearized forms  of \eqn{WHdim} 
and \eqn{polch} and the UV scaling laws obtained by WH--RG and P--RG are 
identical. Then the differentiation of both sides of \eqn{polch} with respect to the 
field variable $\phi$ yields \cite{sg_scheme} 
\beq
\label{derpolch2} 
(2+k\partial_k)\tilde V'_k =  
2\tilde V''_k\tilde V'_k-\frac1{4\pi}\tilde V'''_k 
\eeq 
being independent of the regulator function $K(y)$ and differing of the 
WH--RG equation \eq{derwh} by the term $-\tilde V''_k k\partial_k\tilde V'_k$
and by the opposite sign for the non-linear term. Let us note, that the P--RG 
method treats all quantum fluctuations below and above the scale $k$ on the 
same footing due to the usage of the smooth cutoff. Therefore, even if there 
occurs a scale $k_{\mr{SI}}$ at which $1+\tV_k''$ exhibits zeros, no singular 
behavior is expected in case of the P--RG equation, consequently
Eq.\eq{derpolch2} can be applied above ($k>k_{\mr{SI}}$) and below 
($k<k_{\mr{SI}}$) the scale $k_{\mr{SI}}$ with the price of the SI being 
unnoticed.

\section{Symmetries and Renormalization} 
\label{symmetry}
As a rule, the solution of the RG equations is sought for in a restricted functional 
subspace \cite{rg}. Since the RG equations retain the symmetries of the bare action, 
the functional subspace should be chosen keeping the symmetries of the bare action 
unbroken. Furthermore, even this -- generally infinite dimensional -- subspace is 
reduced to a finite dimensional one by the truncation of the appropriate series 
expansion of the blocked potential. For example, the potential can be expanded 
in powers of the field variable $V_{k}(\phi) = \sum_{n=1}^N \, c_{n}(k) \, \phi^{n}$ 
with a truncation at the power $N$ and the scale-dependence is encoded in the 
coupling constants $c_{n}(k)$. In this case one has to check whether the results 
obtained are independent of $N$. It is known that $O(M)$ scalar models can be 
considered in Taylor expanded form only if $M>1$ (for $M=1$, strong oscillatory 
behavior of the critical exponents in terms of $N$ is observed) \cite{rg}. Similarly, the 
truncated Fourier expanded form can be a straightforward approximation for 
scalar models with periodicity in internal space \cite{sg,sg_prl}.

Let us now turn to the symmetries of MFSG models if the ratios of the frequencies 
are rational. Then the bare potential is periodic in the internal space, let be its 
period $2\pi \beta$, and one has to look for the solution of the RG equations 
among the periodic functions with such a period. The bare potential may have 
however further symmetries as well. For example, the MFSG models can exhibit 
a reflection symmetry besides periodicity. Three cases can be distinguished.

\begin{itemize}
\item 
Let us suppose that the bare potential of the MFSG model contains a single 
cosine mode with $\delta_1 =0$
\beq 
\label{barecos}
\tilde V_{\Lambda}(\phi) = \tilde\mu_1 \, \cos\left(\beta \,\phi\right). 
\eeq 
In this case the model has a discrete reflection symmetry ($\phi \to -\phi$),
which is preserved by the WH--RG and P--RG equations. Since the RG 
transformations generate higher harmonics, one is inclined to look for the 
solution in its Fourier decomposed form 
\beq 
\label{cos} 
\tilde V_{k}(\phi) = \sum_{n=0}^N \, \,  
{\tilde u}_{n}(k) \, \cos\left(n \, \beta \,\phi\right), 
\eeq 
exhibiting periodicity in the internal space. The dimensionless couplings 
are represented by the Fourier amplitudes $\tilde u_{n}(k)$ (with 
$\tilde u_{1}(k=\Lambda) = \tilde\mu_1$) and the `frequency' $\beta$ is a 
scale-independent, dimensionless parameter in the LPA. 
\item
If the bare potential of the MFSG model contains a single sine mode 
(i.e. $\delta_1 = 3\pi/2$)
\beq 
\label{baresin}
\tilde V_{\Lambda}(\phi) = \tilde\mu_1 \, \sin\left(\beta \,\phi\right),
\eeq 
the model has another discrete $Z_2$ symmetry ($\phi \to -\pi/\beta -\phi$)
which is preserved by the RG equations. The potential is antisymmetric 
but the RG equations are not, consequently, one has to look for the solution
of the RG equations as
\begin{align}
\label{sin} 
\tilde V_{k}(\phi) = \sum_{n=0}^N  \left[
{\tilde u}_{2n}(k) \cos\left(2n \beta \phi\right) 
+ {\tilde v}_{2n+1}(k) \sin\left((2n+1) \beta \phi\right)
\right] 
\end{align}
with the dimensionless Fourier amplitudes $\tilde u_{2n}(k)$ and
$\tilde v_{2n+1}(k)$ (and $\tilde v_{1}(k=\Lambda) = \tilde\mu_1$).
Let us note that the double-frequency SG model \eq{dfsg} belongs to
this case, too.
\item
Finally, if the bare potential of the MFSG model contains both cosine
and sine modes  (i.e.  $\delta_1=0$ and  $\delta_2 = 3\pi/2$)
\begin{align}
\label{barecossin}
\tilde V_{\Lambda}(\phi) = \tilde\mu_1 \, \cos\left(\beta \,\phi\right) 
+ \tilde\mu_2 \, \sin\left(\beta \,\phi\right),
\end{align}
the model has no $Z_2$ symmetry, consequently, all the Fourier
modes are generated during the RG flow and the solution has the 
general form 
\begin{align}
\label{cossin} 
\tilde V_{k}(\phi) = \sum_{n=0}^N  \left[
{\tilde u}_{n}(k) \cos\left(n \beta \phi\right) 
+ {\tilde v}_{n}(k) \sin\left(n \beta \phi\right)
\right],
\end{align}
with the dimensionless Fourier amplitudes $\tilde u_{n}(k)$ and
$\tilde v_{n}(k)$ (and $\tilde u_{1}(k=\Lambda) = \tilde\mu_1$,
$\tilde v_{1}(k=\Lambda) = \tilde \mu_2$).
\end{itemize}

Since Eq.\eq{cossin} represents the blocked potential for the most general 
MFSG model with rational frequency ratios, let us further discuss  that case.
Inserting the ansatz \eq{cossin} into the derivative form of the WH--RG equation 
\eq{derwh} one can read off RG flow equations for the Fourier amplitudes, i.e. for 
the scale-dependent dimensionless couplings $\tilde u_n(k)$, $\tilde v_n(k)$ 
which read as
\begin{eqnarray}
\label{whrg1} 
&(2+k\partial_k)n\tilde u_n = \frac{\beta^2}{4\pi}n^3 \tilde u_n + 
\frac{\beta^2}{2}  \sum_{s=1}^N 
\left(s A^{(1)}_{n,s} (2+k\partial_k) \tilde u_s 
+ s A^{(4)}_{n,s} (2+k\partial_k) \tilde v_s
\right),
\\
\label{whrg2}
&(2+k\partial_k)n\tilde v_n = \frac{\beta^2}{4\pi}n^3 \tilde v_n  - 
\frac{\beta^2}{2}  \sum_{s=1}^N 
\left(s A^{(2)}_{n,s} (2+k\partial_k) \tilde u_s 
+ s A^{(3)}_{n,s} (2+k\partial_k) \tilde v_s
\right),
\end{eqnarray}
where 
\begin{align}
A^{(1)}_{n,s}(k) &= (n-s)^2\tu_{|n-s|}-(n+s)^2\tu_{n+s} \Theta(n+s\leq N),
\nn
A^{(2)}_{n,s}(k) &=  {\rm sgn}(s-n) \,\,\, (n-s)^2\tv_{|n-s|} 
+ (n+s)^2 \tv_{n+s} \Theta(n+s\leq N),
\nn
A^{(3)}_{n,s}(k) &= - (n-s)^2\tv_{|n-s|} - (n+s)^2\tv_{n+s} \Theta(n+s\leq N),
\nn
A^{(4)}_{n,s}(k) &=  {\rm sgn}(s-n) \,\,\, (n-s)^2\tv_{|n-s|} - (n+s)^2 \tv_{n+s} \Theta(n+s\leq N),
\nonumber
\end{align}
with ${\rm sgn}(x) = 1$ if $x>0$ and ${\rm sgn}(x) = -1$ if $x<0$, and 
$\Theta(n\leq N) = 1$ if $n \leq N$ and $\Theta(n\leq N) = 0$ if $n > N$.
Let us note that \eqn{derwh} and, consequently, \eqn{whrg1},
\eqn{whrg2} are valid unless SI arises.

Using the same machinery in the framework of the P--RG, one 
obtains from \eqn{derpolch2} the flow equations for $\tilde u_n(k)$
and $\tilde v_n(k)$
\begin{align}
\label{polchrg1} 
&(2+k\partial_k)n\tilde u_n = \frac{\beta^2}{4\pi}n^3 \tilde u_n 
- \beta^2  \sum_{s=1}^N 
\left(s A^{(1)}_{n,s} \tilde u_s + s A^{(4)}_{n,s} \tilde v_s
\right),
\\
\label{polchrg2}
&(2+k\partial_k)n\tilde v_n = \frac{\beta^2}{4\pi}n^3 \tilde v_n  
+ \beta^2  \sum_{s=1}^N 
\left(s A^{(2)}_{n,s} \tilde u_s + s A^{(3)}_{n,s} \tilde v_s
\right),
\end{align}
where $A^{(i)}_{n,s}$, ($i=1,2,3,4$) are the same as those obtained for 
the WH--RG equation. Let us note, that the P--RG method does not take 
into account SI, consequently, \eqn{polchrg1} and \eqn{polchrg2}
are valid at all scales $k$.

Let us end this section with the remark that the strong reduction of the
functional subspace, in particular the truncation of the expansion of the 
blocked potential  in a series of base functions may become unreliable 
when the blocked action becomes almost degenerate, i.e. $1+\tV_k''$ 
approaches zero. This motivates a direct numerical solution of the RG  
equation for  the blocked potential which avoids any assumption on the 
functional subspace where the solution is sought for and any truncated 
series expansion in some base functions \cite{rg,solving_rg,qc}. 
Therefore, we solved the RG equations \eq{derwh} and \eq{derpolch2} 
directly by using a computer algebraic program with periodic boundary 
conditions and the bare initial potential was chosen as a harmonic 
function.

\section{UV scaling} 
\label{uv}
Before the study of the low-energy/IR behavior of the MFSG model, let 
us first discuss the high-energy/UV scaling. This can be achieved by 
the linearization of the RG equations \eq{whrg1}, \eq{whrg2} and 
\eq{polchrg1}, \eq{polchrg2} around the UV Gaussian fixed point 
($\tilde V^*(\phi) \equiv 0$) which results in the following uncoupled 
set of differential equations 
\begin{align}
\label{uv1} 
&(2+k\partial_k) \tilde u_n = \frac{\beta^2}{4\pi} n^2 \tilde u_n, 
\\
\label{uv2}
&(2+k\partial_k) \tilde v_n = \frac{\beta^2}{4\pi} n^2 \tilde v_n,  
\end{align}
which is independent of the RG method used. The solution 
\begin{align}
\label{soluv1} 
&\tilde u_n(k) = \tilde u_n(\Lambda)  
\left(\frac{k}{\Lambda}\right) ^{\frac{\beta^2}{4\pi} n^2 -2}, 
\\
\label{soluv2}
&\tilde v_n(k) = \tilde v_n(\Lambda)  
\left(\frac{k}{\Lambda}\right) ^{\frac{\beta^2}{4\pi} n^2 -2}, 
\end{align}
gives the same UV scaling as that  obtained for the compact MFSG model  
in \cite{rg_multi_sg}. These UV scaling laws can be understood if one considers 
the MFSG model as a perturbation of the corresponding CFT, c.f. the discussion 
below \eqn{conform}. One should conclude that the kinetic term of the action 
suppresses the large amplitude $(\phi^2\gg 1/p^2)$ quantum fluctuations with 
large momentum $(\Lambda^2>p^2>k^2)$ close to the UV cutoff and, therefore 
the UV scaling laws are not influenced by the compactness of the field variable.

\section{IR scaling} 
\label{ir}
In the IR domain neither the kinetic term nor the periodic potential terms are able 
to suppress  the contributions of the large-amplitude quantum fluctuations with
small momenta. Therefore, the compact and non-compact MFSG models are 
expected to behave differently in the IR domain. There are two ways to determine 
the IR scaling of the non-compact MFSG model in the LPA,
(i) either the partial differential equations  \eq{derwh} and \eq{derpolch2} have to 
be solved numerically by a computer algebraic program using the initial condition
\eq{mfsg},
(ii) or one can find the solution of the ordinary differential equations \eq{whrg1}, 
\eq{whrg2} and \eq{polchrg1}, \eq{polchrg2} which are obtained by inserting the 
ansatz \eq{cossin} into Eqs.\eq{derwh}, \eq{derpolch2}. In the latter case, besides 
the LPA, we use a further approximation, namely the truncation of the Fourier 
expansion of the potential.

According to our experiences concerning the renormalization of SG type models 
based on previous publications \cite{sg,qed2_qcd2,qed2,NaEtAl2007,sg_scheme}, 
it is expected that the RG equations obtained by using the truncated Fourier 
decomposition of the periodic potential, is always applicable, except the situation 
if one would like to decide unambiguously whether SI appears or not in the RG flow.
SI is related to the singularity of the RG flow, consequently, in some cases it could 
be important to solve the partial differential (RG) equations without using any further 
approximations in order to be able to decide whether SI can be avoided or not. 
Since the P--RG method does not take account for SI it is more convenient to use 
this method first to consider the IR behavior of the non-compact MFSG model.

\subsection{Polchinski RG approach} 
Let us first discuss the IR effective theory of the MFSG model in the framework 
of the P--RG method by solving  Eqs. \eq{polchrg1},  \eq{polchrg2} numerically
with the most general ansatz \eq{cossin}. 

Qualitatively  different IR scaling behaviors of the MFSG model are observed 
below and above $\beta_c^2 = 8\pi$. For $\beta^2 >8\pi$, every  Fourier amplitudes 
are found to be irrelevant in the limit $k\to 0$, i.e. they are decreasing coupling 
constants independently of the initial conditions, see \fig{polch12pi}. Consequently, 
for $\beta^2 > 8\pi$, the non-compact MFSG model is a free massless theory in 
the IR limit independently of whether the bare initial potential possesses a 
$Z_2$ symmetry (Eqs. \eq{barecos}, \eq{baresin}) or not (Eq.\eq{barecossin}).
\FIGURE{
\epsfig{file=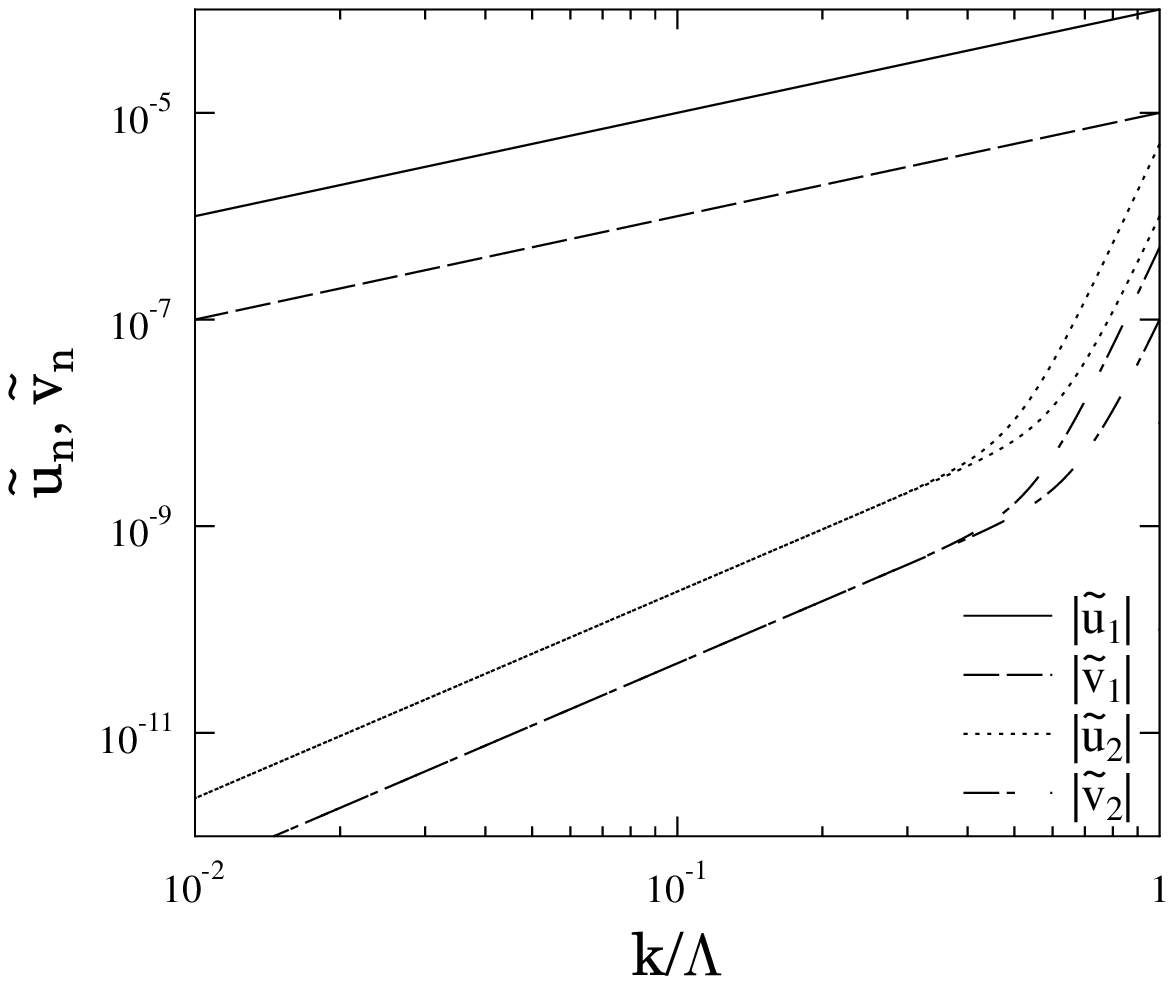,width=10cm} 
\caption{The scaling of the first few Fourier amplitudes of the non-compact 
MFSG model is obtained by the P--RG method solving  Eqs. \eq{polchrg1}, 
\eq{polchrg2} numerically for $\beta^2 =12\pi$ with various initial conditions 
for the higher harmonics.} 
\label{polch12pi}
} 

A further important result of the RG analysis is that the Fourier amplitudes of 
the non-compact MFSG model show up the IR scaling behavior
\begin{align}
\label{solir1} 
&\tilde u_n(k) = f_n
\left(\frac{k}{\Lambda}\right) ^{n (\frac{\beta^2}{4\pi} -2)}, 
\\
\label{solir2}
&\tilde v_n(k) = g_n 
\left(\frac{k}{\Lambda}\right) ^{n (\frac{\beta^2}{4\pi} -2)}, 
\end{align}
which differs from that obtained in the UV regime, Eqs. \eq{soluv1} and 
\eq{soluv2}. Here the well-justified approximations 
\begin{align}
\sum_{s=1}^N s A^{(1)}_{n,s} \tilde u_s &\approx 
+\sum_{s=1}^{n-1} s (n-s)^2 f_{n-s} f_{s}  
\left(\frac{k}{\Lambda}\right)^{n(\frac{\beta^2}{4\pi} -2)},
\nn
\sum_{s=1}^N s A^{(2)}_{n,s} \tilde u_s &\approx 
-\sum_{s=1}^{n-1} s (n-s)^2 g_{n-s} f_{s}  
\left(\frac{k}{\Lambda}\right)^{n(\frac{\beta^2}{4\pi} -2)},
\nn
\sum_{s=1}^N s A^{(3)}_{n,s} \tilde v_s &\approx 
-\sum_{s=1}^{n-1} s (n-s)^2 f_{n-s} g_{s} 
\left(\frac{k}{\Lambda}\right)^{n(\frac{\beta^2}{4\pi} -2)},
\nn
\sum_{s=1}^N s A^{(4)}_{n,s} \tilde v_s  &\approx 
-\sum_{s=1}^{n-1} s (n-s)^2 g_{n-s} g_{s} 
\left(\frac{k}{\Lambda}\right)^{n(\frac{\beta^2}{4\pi} -2)},
\nonumber
\end{align}
result in the following recursion relations for the constants
$f_n$ and $g_n$,
\begin{align}
\label{irpolch1}
&f_n = 
-\frac{\beta^2  \sum_{s=1}^{n-1} s (n-s)^2 (f_{n-s} f_s - g_{n-s} g_s)}
{n\left[2+ n\left (\frac{\beta^2}{4\pi} -2\right)-\frac{\beta^2}{4\pi}n^2\right]},
\\
\label{irpolch2}
&g_n = 
+\frac{\beta^2  \sum_{s=1}^{n-1} s (n-s)^2 (g_{n-s} f_s + f_{n-s} g_s)}
{n\left[2+ n\left (\frac{\beta^2}{4\pi} -2\right)-\frac{\beta^2}{4\pi}n^2\right]}.
\end{align}
Let us note that in case of the fundamental modes (i.e. for $n=1$) the UV 
and IR scalings coincide. The IR scaling of the model is 
determined by two independent parameters, $f_1 = \tilde u_1(\Lambda)$  
and $g_1 = \tilde v_1(\Lambda)$ since for $n>1$ the constants $f_n$ 
and $g_n$ are fixed by Eqs.\eq{irpolch1}, \eq{irpolch2}. Therefore, the IR 
behavior of the model is independent of the initial conditions for the higher 
harmonics, see \fig{polch12pi}, and depends on either $\tilde u_1(\Lambda)$
or $\tilde v_1(\Lambda)$ if the model has a $Z_2$ symmetry and in the absence 
of the reflection symmetry the IR physics is determined by both  
$\tilde u_1(\Lambda)$ and $\tilde v_1(\Lambda)$.

For $\beta^2 <8\pi$, one has to distinguish three scaling regimes in case of the 
non-compact MFSG model (i) the UV (ii) the IR (iii) and the deep IR scaling 
behavior, see \fig{polch4pi}. The UV \eq{soluv1}, \eq{soluv2} and the IR \eq{solir1}, 
\eq{solir2} scaling laws are given by the same expressions as those obtained 
in the strong coupling phase ($\beta^2>8\pi$). However, if  $\beta^2 <8\pi$, 
according to the IR scaling law, every Fourier amplitude becomes relevant 
(increasing) coupling in the IR domain. Even more important difference is that a 
qualitatively new behavior is  found in the deep IR limit ($k\to 0$), namely, at a 
certain momentum scale $k_c$ the Fourier amplitudes of the non-compact MFSG 
model become constants, see \fig{polch4pi}. 
\FIGURE{
\epsfig{file=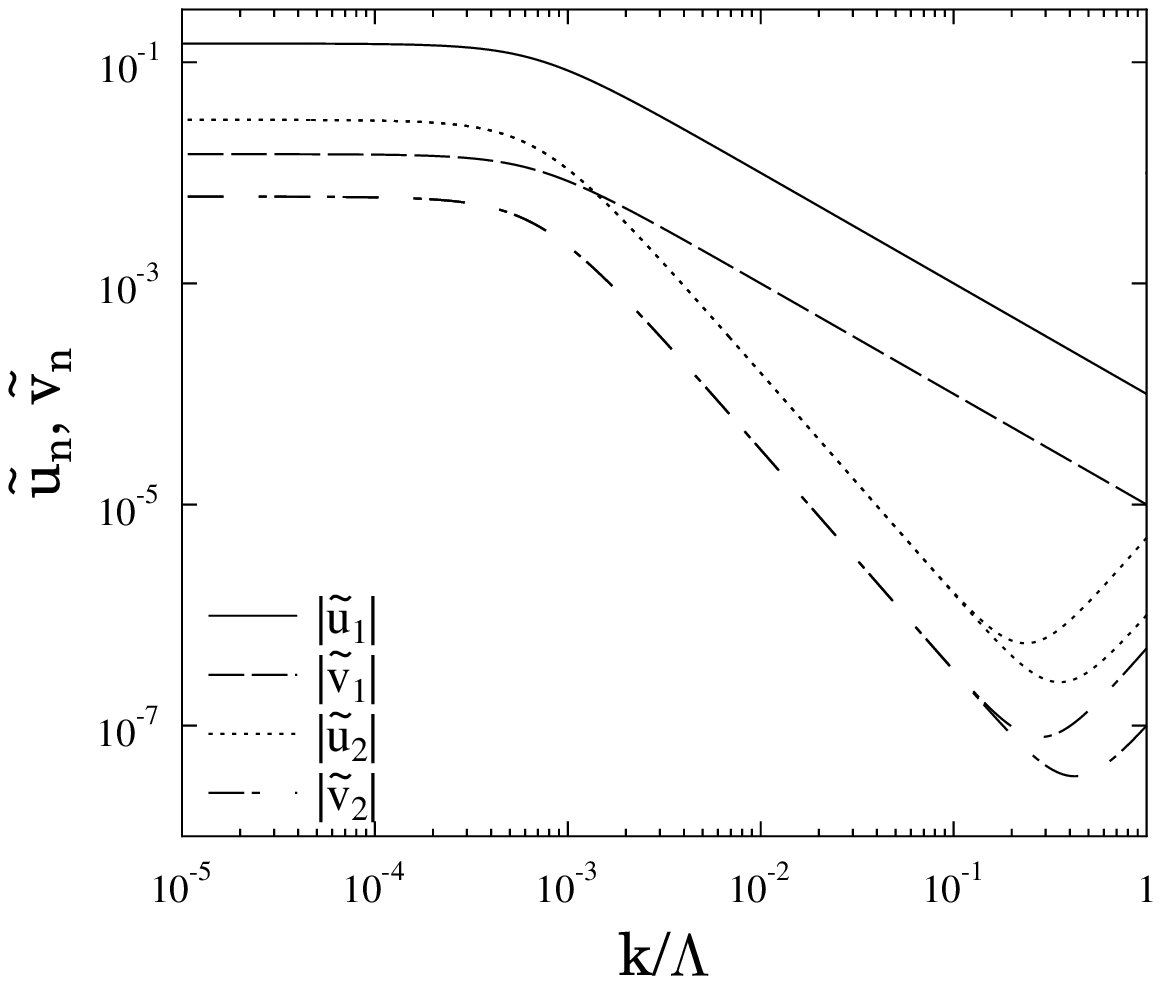,width=10cm} 
\caption{The scaling of the first few Fourier amplitudes of the non-compact 
MFSG model is obtained by the P--RG method solving  Eqs. \eq{polchrg1}, 
\eq{polchrg2} numerically for $\beta^2 =4\pi$  with various initial conditions 
for the higher harmonics. At the momentum scale $k_c \sim 3\times 10^{-4}$, 
the Fourier amplitudes become constants.} 
\label{polch4pi}
}
Therefore, if $\beta^2 <8\pi$, the dimensionless IR effective potential of the 
non-compact model is non-trivial. 

Let us analyze the sensitivity of the IR
theory on the initial conditions in order to map out the phase structure.  
If the bare action has no $Z_2$ symmetry (see Eq.\eq{barecossin}) then the 
deep IR effective potential depends on a single parameter, namely, the ratio 
of the initial values of the fundamental modes, 
$r=\tilde u_1(\Lambda)/\tilde v_1(\Lambda)$ which remains unchanged 
during the RG flow, see \fig{polch4pisens}.
Let us remind that in the strong coupling regime ($\beta^2 >8\pi$) 
the IR behavior of the model (without a $Z_2$ symmetry) is determined by 
two independent parameters ($\tilde u_1(\Lambda)$ and $\tilde v_1(\Lambda)$).
In the presence of  $Z_2$  symmetry (see \eq{barecos} for $r=\infty$ or
\eq{baresin} for $r=0$), the deep IR potential is superuniversal, i.e. it is 
independent of any initial conditions \cite{sg,sg_scheme}.

Again, for $\beta^2 >8\pi$, if the action has a reflection symmetry, the IR scaling
is determined by a single parameter, i.e. the initial value of the fundamental 
mode, (either $\tilde u_1(\Lambda)$ or $\tilde v_1(\Lambda)$).
\FIGURE{
\epsfig{file=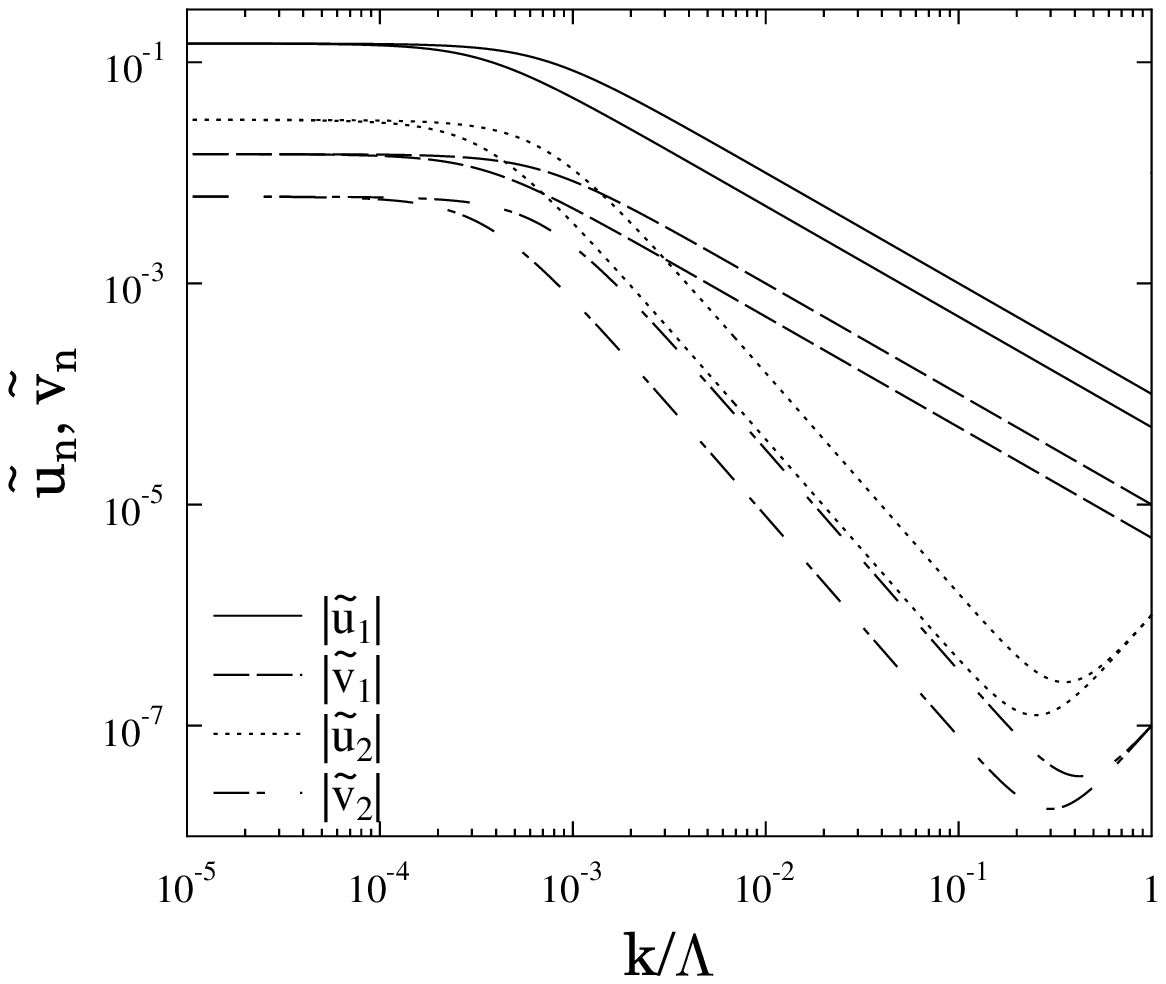,width=10cm} 
\caption{In this figure we show that the IR effective potential of the 
non-compact MFSG model for $\beta^2=4\pi$ depends on only the ratio
of the Fourier amplitudes of the fundamental cosine and sine modes. If 
the P--RG equation has been solved with various initial conditions for 
$\tilde u_1(\Lambda)$ and $\tilde v_1(\Lambda)$ but keeping their ratio
fixed, then one obtains the same deep IR behavior.}
\label{polch4pisens} 
} 
Therefore, the non-compact MFSG model has two phases separated by the 
critical value $\beta_c^2=8\pi$. As a consequence of the superuniversal,
and universal behavior, no other phase transition can be identified in the 
non-compact model.  

Finally let us consider the IR behavior of the non-compact MFSG model by
solving directly the P--RG equation \eq{derpolch2}. When the solution of the 
partial differential equation \eq{derpolch2} had been obtained it was expanded 
in Fourier series. For $\beta^2=4\pi$ the UV, IR and the deep IR scaling of the 
first few Fourier amplitudes coincide to that of obtained by the numerical solution 
of Eqs. \eq{polchrg1},  \eq{polchrg2} which are plotted in \fig{polch4pi}. 
Therefore, there is an excellent quantitative agreement between the results 
obtained by solving Eqs. \eq{polchrg1}, \eq{polchrg2} and by solving 
Eq.\eq{derpolch2} directly. This shows that in case of the P--RG method 
the RG flow seems to avoid the SI and 
one can look for the solution of the RG equations in its Fourier decomposed form.

If one tries to determine the IR behavior of the MFSG model by an RG method,
like the WH--RG approach, which has a singular structure, and consequently,
SI could appear in the RG flow it could be important to solve the partial differential 
RG equation obtained in the LPA without using any further approximations in 
order to be able to decide whether SI can be avoided or not.

\subsection{Wegner-Houghton RG approach} 
Let us consider the IR effective theory of the MFSG model in the framework 
of the WH--RG method by solving  Eqs. \eq{whrg1},  \eq{whrg2} numerically.
For $\beta^2 >8\pi$, similarly to the results obtained by the P--RG method, 
the Fourier amplitudes are irrelevant in the limit $k\to 0$, independently of 
the initial conditions, see \fig{wh12pi}. 
\FIGURE{
\epsfig{file=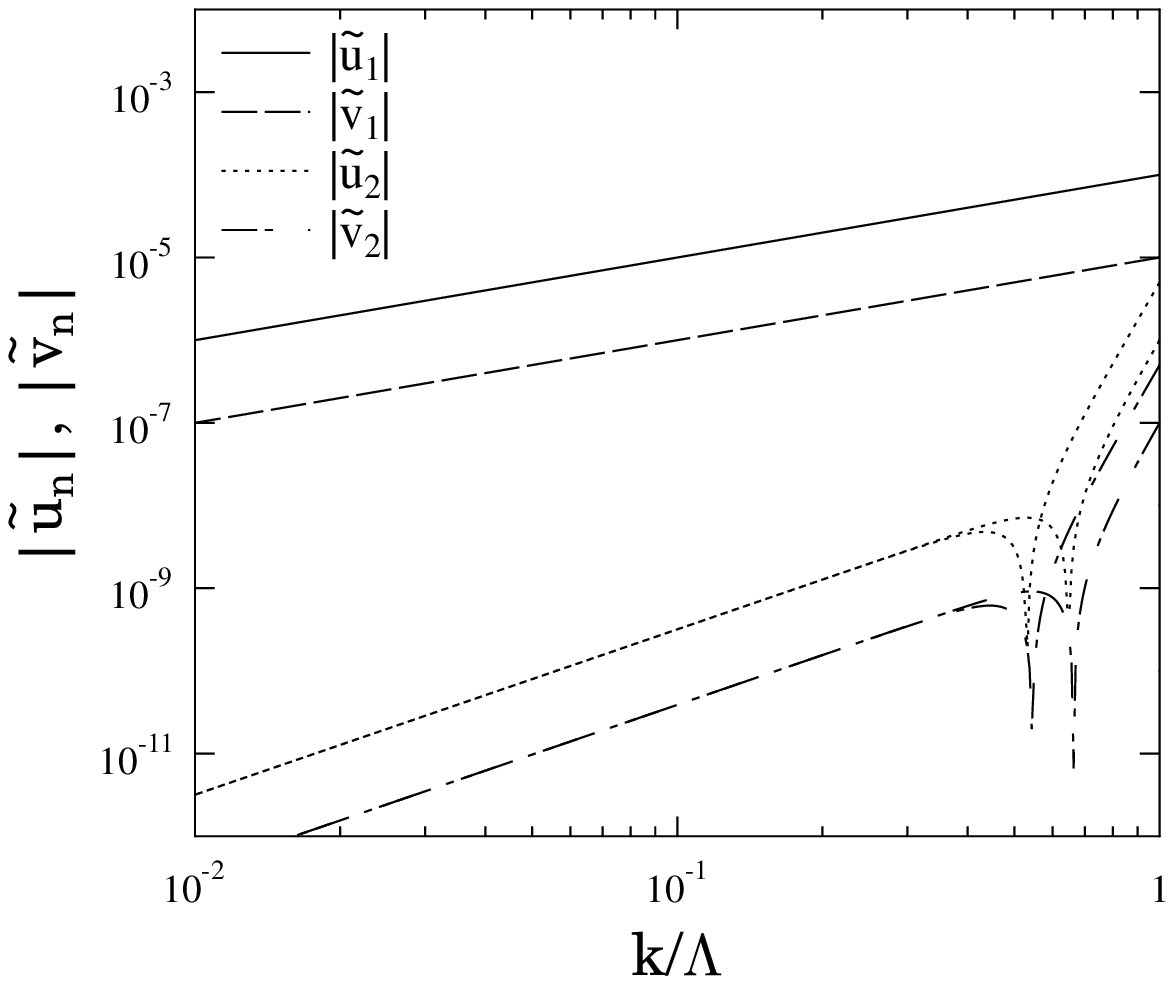,width=10cm} 
\caption{The scaling of the first few Fourier amplitudes of the non-compact 
MFSG model is obtained by the WH--RG method solving  Eqs. \eq{whrg1}, 
\eq{whrg2} numerically for $\beta^2 =12\pi$ with various initial conditions 
for the higher harmonics. The peaks in the scaling of $\tilde u_2(k)$ and
$\tilde v_2(k)$ indicate the change of their sign during the RG flow.}
\label{wh12pi}
} 
The numerical solution of the  WH--RG equation provides once again the 
IR scaling laws given by \eq{solir1} and \eq{solir2}. Similarly to the P--RG 
flow, this IR behavior can also be obtained by using the IR approximations
\begin{align}
&\sum_{s=1}^N s A^{(1)}_{n,s} (2+k\partial_k)  \tilde u_s \approx 
+\sum_{s=1}^{n-1} s (n-s)^2 f_{n-s} f_{s}  
\left[2+s \left(\frac{\beta^2}{4\pi} -2\right)\right]
\left(\frac{k}{\Lambda}\right)^{n(\frac{\beta^2}{4\pi} -2)},
\nn
&\sum_{s=1}^N s A^{(2)}_{n,s}  (2+k\partial_k) \tilde u_s \approx 
-\sum_{s=1}^{n-1} s (n-s)^2 g_{n-s} f_{s}  
\left[2+s \left(\frac{\beta^2}{4\pi} -2\right)\right]
\left(\frac{k}{\Lambda}\right)^{n(\frac{\beta^2}{4\pi} -2)},
\nn
&\sum_{s=1}^N s A^{(3)}_{n,s}  (2+k\partial_k) \tilde v_s \approx 
-\sum_{s=1}^{n-1} s (n-s)^2 f_{n-s} g_{s} 
\left[2+s \left(\frac{\beta^2}{4\pi} -2\right)\right]
\left(\frac{k}{\Lambda}\right)^{n(\frac{\beta^2}{4\pi} -2)},
\nn
&\sum_{s=1}^N s A^{(4)}_{n,s}  (2+k\partial_k) \tilde v_s  \approx 
-\sum_{s=1}^{n-1} s (n-s)^2 g_{n-s} g_{s}
\left[2+s \left(\frac{\beta^2}{4\pi} -2\right)\right]
\left(\frac{k}{\Lambda}\right)^{n(\frac{\beta^2}{4\pi} -2)},
\nonumber
\end{align}
which result in the recursion relations for $f_n$ and $g_n$,
\begin{align}
\label{irwh1}
f_n = +\frac{\beta^2 \sum_{s=1}^{n-1} s (n-s)^2 (f_{n-s} f_s - g_{n-s} g_s)[1-s+\frac{s\beta^2}{8\pi}]}
{n\left[2+n \left(\frac{\beta^2}{4\pi} -2\right) -\frac{\beta^2}{4\pi}n^2\right]},
\\
\label{irwh2}
g_n = -\frac{\beta^2 \sum_{s=1}^{n-1} s (n-s)^2 (g_{n-s} f_s + f_{n-s} g_s)[1-s+\frac{s\beta^2}{8\pi}]}
{n\left[2+n \left(\frac{\beta^2}{4\pi} -2\right) -\frac{\beta^2}{4\pi}n^2\right]}.
\end{align}
This shows that the IR scalings of the non-compact MFSG model 
determined by the P--RG and the WH--RG methods are qualitatively the same
for $\beta^2>8\pi$. The UV/IR scalings of the fundamental modes (i.e. for $n=1$) 
coincide, independently of the RG method used, $f_1 = \tilde u_1(\Lambda)$  
and $g_1 = \tilde v_1(\Lambda)$. The IR constants $f_n$ and $g_n$ of the 
higher harmonics (i.e. $n>1$) are determined by the equations \eq{irwh1}, 
\eq{irwh2} which predict the IR behavior similar to that obtained by the P--RG 
method. However, in case of the WH--RG method the $f_n$, $g_n$ parameters 
have alternating signs for even and odd values of $n$. Therefore, if we use 
the same initial conditions, (e.g. all the bare Fourier amplitudes are positive) 
then in case of the WH--RG method, $\tilde u_2(k)$ and $\tilde v_2(k)$ change 
their signs during the RG flow, see  \fig{wh12pi}. It is important to note that the 
IR scaling of the model (similarly to the P--RG method) is determined by two 
independent parameters ($\tilde u_1(\Lambda)$, $\tilde v_1(\Lambda)$), if the 
bare action has no $Z_2$ symmetry and depends on a single parameter (either 
$\tilde u_1(\Lambda)$ or $\tilde v_1(\Lambda)$) in case of a $Z_2$ symmetric 
bare action, and it is independent of the initial conditions of the higher harmonics, 
see \fig{wh12pi}. In conclusion, the WH--RG and the P--RG methods produce the 
same IR behavior for the MFSG model if $\beta^2 >8\pi$.

For $\beta^2 <8\pi$, the IR scaling behavior turns all the Fourier amplitudes into 
relevant coupling constants, consequently, the logarithm of the WH--RG equation 
\eq{WHdim} could become infinite, hence a SI could appear in the WH--RG flow.
Indeed, in \fig{wh4pi} the scaling of the coupling constants of the non-compact
MFSG model is presented for $\beta^2 = 4\pi$ and the vertical line shows the 
appearance of SI. 
\FIGURE{
\epsfig{file=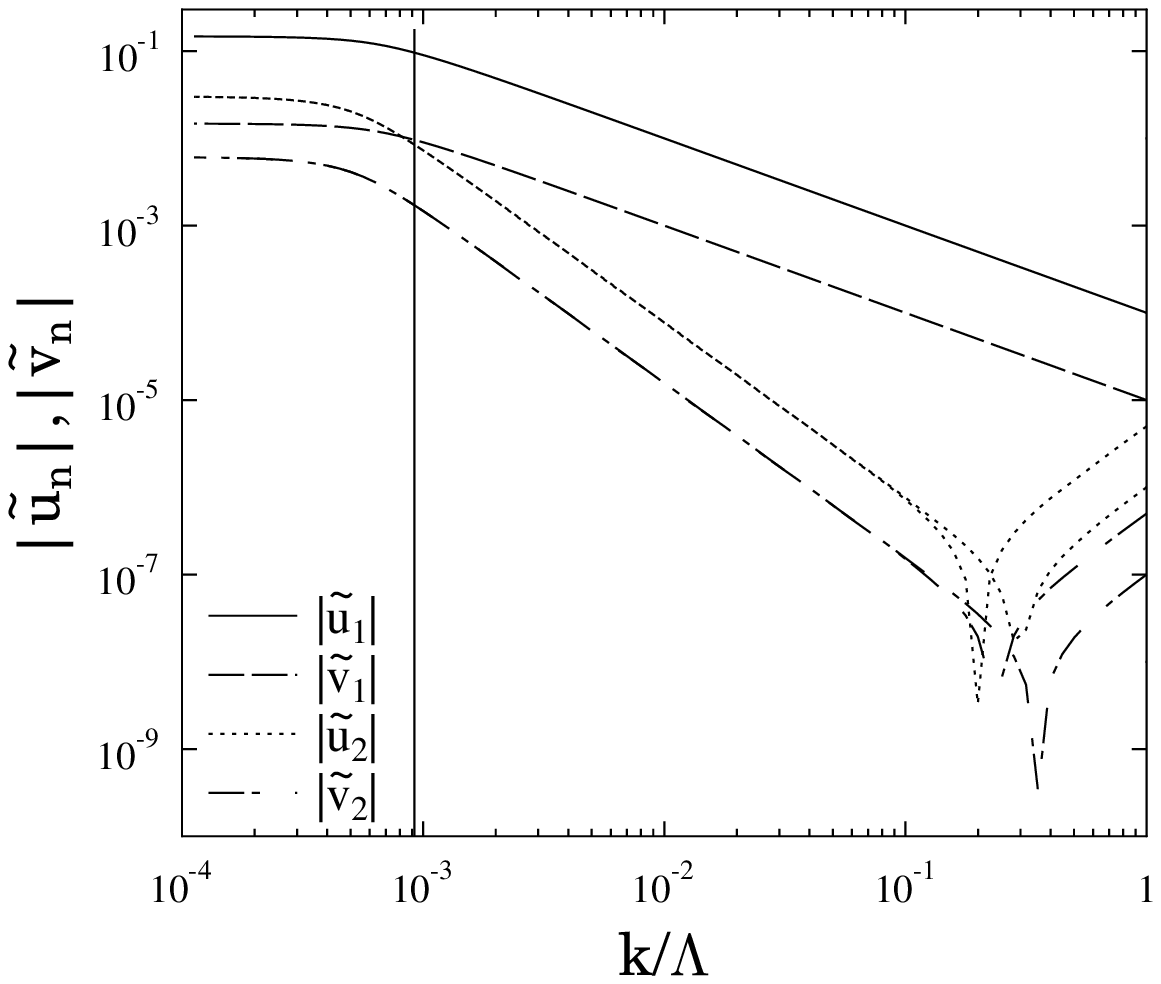,width=10cm} 
\caption{The scaling of the first few Fourier amplitudes of the non-compact 
MFSG model is obtained in the framework of the WH--RG method for 
$\beta^2 =4\pi$ by solving numerically either Eqs. \eq{whrg1}, \eq{whrg2}
or Eq. \eq{derwh}. In the latter case, the partial differential equation \eq{derwh} 
is solved by a computer algebraic code and then the solution is expanded in Fourier 
series. The vertical line indicates the momentum scale of SI ($k_{\rm SI}$) where 
Eqs. \eq{whrg1}, \eq{whrg2} lose their validity. Above this scale, $k_{\rm SI}<k$, 
the results obtained by Eqs. \eq{whrg1}, \eq{whrg2} and by Eq. \eq{derwh} coincide. 
Below the scale of SI, $k<k_{\rm SI}$, the scaling of the Fourier amplitudes is 
determined by the direct integration of Eq. \eq{derwh}.} 
\label{wh4pi}
} 
Beyond the momentum scale $k_{\rm SI}$, the WH--RG 
equation loses its validity and one has to use the tree-level RG equation 
\eq{treedim} which leads to the IR effective potential \eq{treewhpot} in the deep 
IR limit ($k\to 0$). In order to preserve periodicity, the IR effective potential of the
MFSG model has a parabola-shape for $\phi\in [-\pi/\beta,\pi/\beta]$ and such 
parabola sections are repeated along the $\phi$ axis. Let us analyze the sensitivity
of the IR effective theory on the UV initial conditions. In case of a reflection 
symmetry $\phi \to -\phi$, the linear term vanishes in \eq{treewhpot}, i.e. $c=0$, 
and the potential is superuniversal, i.e. independent of  any initial conditions.
If the bare action has another type of reflection symmetry $\phi \to -\phi -\pi/\beta$,
then the constant in \eq{treewhpot} is non-zero but fixed, i.e. $c= -\pi/2\beta$, 
consequently, the IR potential is again superuniversal. If the bare action of the 
MFSG model has no $Z_2$ symmetry then the deep IR behavior depends on 
a single parameter $c$. Therefore, in the framework of the WH--RG method if SI 
appears in the RG flow, the sensitivity of the IR behavior on the UV parameters 
is found to be the same as that  obtained by the P--RG approach. 

Finally, let us consider the IR scaling of the non-compact MFSG model by
solving the WH--RG equation \eq{derwh} by a computer algebraic code. The 
solution found is expanded in Fourier series in order to compare the results to 
those obtained by Eqs. \eq{whrg1}, \eq{whrg2}. For $\beta^2=4\pi$ the scalings 
of the first few Fourier amplitudes are plotted in \fig{wh4pi}. There is a quantitative 
agreement between the results obtained by Eq.\eq{derwh} and Eqs. \eq{whrg1}, 
\eq{whrg2} in the UV and IR scaling regimes. However, the important difference is 
that no SI is found in the RG flow when Eq.\eq{derwh} is solved directly. This 
indicates that SI occurs in the WH--RG approach as an artifact due to the truncated 
Fourier-expansion applied to the almost degenerate blocked action of the MFSG 
model, at least  for $\beta^2 = 4\pi$. On the other hand, it seems to support the 
Quantum Censorship conjecture to be at work in the MFSG model as well \cite{qc}. 
Let us emphasize that independently of whether the blocked action becomes 
degenerate or not, the sensitivity of the deep IR behavior of the MFSG model on 
the UV initial parameters is found to be the same. Consequently, the phase structure 
of the non-compact MFSG model is determined unambiguously and independently 
of the RG method used.

Let us note that if Quantum Censorship is really on work, then the WH--RG 
\eq{derwh} and P--RG \eq{derpolch2} partial differential equations and also 
their Fourier expanded forms, Eqs. \eq{whrg1}, \eq{whrg2} and Eqs. \eq{polchrg1}, 
\eq{polchrg2} retain their validity in the deep IR regime. When there the Fourier 
amplitudes take constant values at some momentum scale $k_c$, 
i.e. $\partial_k \tilde V_{(k<k_c) }=0$ or $\partial_k \tilde u_n(k<k_c) =0$, 
$\partial_k \tilde v_n(k<k_c) =0$ hold, then Eqs. \eq{whrg1}, \eq{whrg2} and 
Eqs. \eq{polchrg1}, \eq{polchrg2} reduce to the same recursion equations except 
the sign of the non-linear term. Consequently, in the IR limit $k\to0$ the WH--RG 
and the P--RG methods result in the same absolute values of the couplings
$|\tilde u_n(0)|$ and $|\tilde v_n(0)|$ of the MFSG model.

\section{Summary} 
\label{sum}
In this paper we considered how the compactness of the field influences  
the renormalization and, consequently, the low-energy behavior of the theory. 
In particular, we compared the high-energy/UV and low-energy/IR behaviors 
of the two-dimensional multi-frequency sine--Gordon (MFSG) scalar field model 
defined by compact and non-compact field variables. We studied 
the renormalization of the MFSG model with a non-compact field in the 
framework of the functional renormalization group (RG) method using the local 
potential approximation (LPA), discussing the comparison with the results 
for the compact double- and multi- frequency sine Gordon. 

We showed  that the UV scaling of the compact and the non-compact MFSG 
models coincides but their IR behaviors are different. In the UV limit, the 
quantum fluctuations (with high frequency and small amplitude) do not feel the 
difference between the models defined by compact and non-compact fields but 
different behaviours are expected to appear in the IR limit due to the large-amplitude 
quantum fluctuations of the IR domain. On the one hand the critical frequency 
$\beta^2_c = 8\pi$ at which the sine-Gordon model undergoes a topological 
phase transition is found to be unaffected by the compactness of the field since 
it is determined by the UV scaling laws. On the other hand, while it is known that 
the compact model has first and second order (Ising) type phase transitions which 
are determined by the IR scaling, we showed that these are absent in the 
non-compact model.

Indeed, the IR effective potential of the non-compact MFSG model was found to be 
different above and below $\beta^2_c = 8\pi$. For $\beta^2> 8\pi$, the deep IR 
behavior of the non-compact MFSG model with $Z_2$ symmetry (i.e. $\phi \to -\phi$
or $\phi \to -\pi/\beta -\phi$) depends on the UV initial condition for either the 
fundamental cosine or the fundamental sine mode, respectively, and for 
$\beta^2 < 8\pi$ it is superuniversal, i.e. independent of any initial conditions. If the
non-compact MFSG model has no $Z_2$ symmetry, for $\beta^2> 8\pi$ the IR 
effective potential depends on the UV initial conditions both for the fundamental 
cosine and sine modes (i.e. it depends on two independent parameters) and for 
$\beta^2 < 8\pi$ it is universal, i.e. depends on only a single parameter, namely 
the ratio $\tilde u_1(\Lambda)/\tilde v_1(\Lambda)$. Consequently, due to the 
superuniversal and universal IR behavior of the non-compact MFSG model, there 
is no room for first or second order phase transitions for $\beta^2 < 8\pi$.

These results were obtained by the functional renormalization group analysis of 
the non-compact MFSG model in the framework of the Polchinski and the 
Wegner--Houghton RG methods where the latter is mathematically equivalent to 
the effective average action RG with the power-law regulator ($b=1$) and the 
functional Callan-Symanzik RG equation. The RG flow of the non-compact MFSG 
model was determined in two different ways (i) either the RG equations obtained 
in the LPA were solved numerically by a computer algebraic code and then the 
solution expanded in Fourier series, (ii) or first the RG equations were derived for 
the Fourier amplitudes and then those solved numerically. In the latter case, it was 
unavoidable to implement a further approximation besides the LPA, namely the 
truncation of the Fourier expansion of the potential.
The sensitivity of the IR effective potential on the UV initial conditions, and 
consequently, the phase structure was found to be the same in both cases. 
Moreover, except the situation where the RG flow has a singularity, i.e. a spinodal
instability (SI) appears in the IR limit, the scaling of the Fourier amplitudes obtained 
in the above mentioned two different ways, coincide. For $\beta^2 < 8\pi$, in case 
of the non-compact MFSG model a momentum scale was generated by the RG
transformation in the deep IR regime (either the scale where the Fourier amplitudes
become constants or the scale of SI). Below this momentum scale, the theory 
becomes superuniversal (if it has a $Z_2$ symmetry) or universal (if it has no
$Z_2$ symmetry).  

Finally, the classification of the IR scaling operators into relevant, 
marginal or irrelevant ones was also found to be different in case of the compact 
and the non-compact MFSG models. For the compact model, one can rely on the UV
results even in the IR limit but for the non-compact case new types of scaling laws 
were observed in the IR domain which modify  the classification of the scaling 
operators.

\section*{Acknowledgments} 
Fruitful discussions with G. Delfino, G. Mussardo and P. Sodano are warmly 
acknowledged. I.N., S.N., and K.S. acknowledge the support by the project 
T\'AMOP 4.2.1-08/1-2008-003. The project is implemented through the New 
Hungary Development Plan co-financed by the European Social Fund, and 
the European Regional Development Fund. A.T. acknowledges support 
by the grants INSTANS (from ESF) and 2007JHLPEZ (from MIUR).

 

\end{document}